\documentclass[10pt,letterpaper]{article}

\usepackage{cogsci}
\usepackage{pslatex}
\usepackage{apacite}
\usepackage{graphicx}
\usepackage{algorithm}
\usepackage[noend]{algpseudocode}
\usepackage{subfig}

\title{Converting Cascade-Correlation Neural Nets into Probabilistic Generative Models}
\author{{\large \bf Ardavan S. Nobandegani$^{1,3}$ \quad Thomas R. Shultz$^{2,3}$}\\
\{ardavan.salehinobandegani@mail.mcgill.ca, thomas.shultz@mcgill.ca\}\\
$^{1}$Department of Electrical and Computer Engineering, McGill University\\
$^{2}$School of Computer Science, McGill University\\
$^{3}$Department of Psychology, McGill University}

\newcommand{\bb}[1]{\textbf{#1}}

\newcommand{\mc}[1]{\mathcal{#1}}

\usepackage{amsmath}
\usepackage{amsfonts}
\usepackage{amssymb}
\usepackage{amsthm}
\usepackage{color}
\usepackage{amsmath}
\usepackage{amsfonts}
\usepackage{amssymb}
\usepackage{bm}

\makeatletter
\def\BState{\State\hskip-\ALG@thistlm}
\makeatother

\begin{document}
\maketitle

\begin{abstract}
Humans are not only adept in recognizing what class an input instance belongs to (i.e., classification task), but perhaps more remarkably, they can imagine (i.e., \emph{generate})  plausible instances of a desired class with ease, when prompted. Inspired by this, we propose a framework which allows transforming Cascade-Correlation Neural Networks (CCNNs) into probabilistic generative models, thereby enabling CCNNs to generate samples from a category of interest. CCNNs are a well-known class of deterministic, discriminative NNs, which autonomously construct their topology, and have been successful in giving accounts for a variety of psychological phenomena. Our proposed framework is based on a Markov Chain Monte Carlo (MCMC) method, called the Metropolis-adjusted Langevin algorithm, which capitalizes on the gradient information of the target distribution to direct its explorations towards regions of high probability, thereby achieving good mixing properties. Through extensive simulations, we demonstrate the efficacy of our proposed framework.  
\end{abstract}

\section{Introduction}
A green-striped elephant! No one has probably seen such a thing---no surprise. But what is a surprise is our ability to {imagine} one with almost no trouble. Humans are not only adept in recognizing what class an input instance belongs to (i.e., classification task), but more remarkably, they can imagine (i.e., \emph{generate}) plausible instances of a desired class with ease, when prompted. In fact, humans can generate instances of a desired class, say, {elephant}, that they have never encountered before, like, a green-striped elephant.\footnote{In \emph{counterfactual} terms: Had a human seen a green-striped elephant, s/he would have yet recognized it as an elephant. Geoffrey Hinton once told a similar story about a pink elephant!} In this sense, humans' generative capacity goes beyond merely retrieving from memory. In computational terms, the notion of generating examples from a desired class can be formalized in terms of \emph{sampling} from some underlying {probability distribution}, and has been extensively studied in machine learning under the rubric of {probabilistic generative models}. 

Cascade-Correlation Neural Networks (CCNNs) \cite{fahlman1989cascade} are a well-known class of {discriminative} (as opposed to generative) models that have been successful in simulating a variety of phenomena in the developmental literature, e.g., infant learning of word-stress patterns in artificial languages \cite{shultz2006neural}, syllable boundaries \cite{shultz2006neural}, visual concepts \cite{shultz2006constructive}, and have also been successful in capturing important developmental regularities in a variety of tasks, e.g., the balance-scale task \cite{shultz1994modeling,shultz2007rule}, transitivity \cite{shultz2004connectionist}, conservation \cite{shultz1998computational}, seriation  \cite{mareschal1999development}. Moreover, CCNNs exhibit several similarities with known brain functions:
distributed representation, self-organization of network topology, layered hierarchical topologies, both cascaded and direct pathways, an S-shaped activation function, activation modulation via integration of neural inputs, long-term potentiation, growth at the newer end of the network via synaptogenesis or neurogenesis, pruning, and weight freezing \cite{westermann2006modeling}. Nonetheless, in virtue of being deterministic and discriminative, CCNNs have so far lacked the capacity to probabilistically generate examples from a category of interest. This ability can be used, e.g., to diagnose what the network knows at various points during training, particularly when dealing with high-dimensional input spaces.

In this work, we propose a framework which allows transforming CCNNs into probabilistic generative models, thereby enabling CCNNs to generate samples from a category. Our proposed framework is based on a Markov Chain Monte Carlo (MCMC) method, called the Metropolis-Adjusted Langevin (MAL) algorithm, which employs the {gradient} of the target distribution to guide its explorations towards regions of high probability, thereby significantly reducing the undesirable random walk often observed at the beginning of an MCMC run (a.k.a. the burn-in period). MCMC methods are a family of algorithms for sampling from a desired probability distribution, and have been successful in simulating important aspects of a wide range of cognitive phenomena, e.g., temporal dynamics of multistable perception \cite{gershman2012multistability,moreno2011bayesian}, developmental changes in cognition \cite{bonawitz2014probabilistic}, category learning \cite{sanborn2010rational}, causal reasoning in children \cite{bonawitz2014win}, and giving accounts for many cognitive biases \cite{dasgupta2016hypotheses}.

Furthermore, work in theoretical neuroscience has shed light on  possible mechanisms according to which MCMC methods could be realized in generic cortical circuits \cite{buesing2011neural,moreno2011bayesian,pecevski2011probabilistic,gershman2016complex}. In particular, \citeA{moreno2011bayesian} showed how an attractor neural network implementing MAL could account for multistable perception of drifting gratings, and \citeA{savin2014spatio} showed how a network of leaky integrate-and-fire neurons could implement MAL in a biologically-realistic manner.

\section{Cascade-Correlation Neural Networks}
CCNNs are a special class of deterministic artificial neural networks, which construct their topology in an autonomous fashion---an appealing property simulating developmental phenomena \cite{westermann2006modeling} and other cases where networks need to be constructed. CCNN training starts with a two-layer network (i.e., the input  and the output layer) with no hidden units, and proceeds by recruiting hidden units one at a time, as needed. Each new hidden unit is trained to be maximally correlated with residual error in the network built so far, and is recruited into a hidden layer of its own, giving rise to a deep network with as many hidden layers as the number of recruited hidden units. CCNNs use sum-of-squared error as objective function, and typically use symmetric sigmoidal activation functions with range $-0.5$ to $+0.5$ for hidden and output units.\footnote{\citeA{fahlman1989cascade} also suggest linear, Gaussian, and asymmetric sigmoidal (with range $0$ to $+1$) activation functions as alternatives. Our proposed framework can be straightforwardly adapted to handle all such activation functions.} Some variants have been proposed for CCNNs, e.g., Sibling-Descendant Cascade-Correlation (SDCC) \cite{baluja1994reducing} and Knowledge-Based Cascade-Correlation (KBCC) \cite{shultz2001knowledge}. Although in this work we specifically focus on standard CCNNs, our proposed framework can handle SDCC and KBCC as well.

\section{The Metropolis-Adjusted Langevin Algorithm}
\label{sec_MAL_alg}
MAL \cite{roberts1996exponential} is a special type of MCMC method, which employs the {gradient} of the target distribution to guide its explorations towards regions of high probability, thereby reducing the burn-in period. More specifically, MAL combines the two concepts of Langevin dynamics (a random walk guided by the gradient of the target distribution), and the Metropolis-Hastings algorithm (an accept/reject mechanism for generating a sequence of samples the distribution of which asymptotically converges to the target distribution).
\begin{algorithm}
\caption{{The Metropolis-Adjusted Langevin algorithm}}\label{MAL_alg}
\begin{algorithmic}[1]
\Statex \textbf{Input}: Target distribution $\pi$(\bb X), parameter $\tau\in\mathbb{R}_+$, number of samples $N$.
\Statex \textbf{Output}: Samples $\bb X^{(0)},\ldots,\bb X^{(N-1)}$.
\State Pick $\bb X^{(0)}$ arbitrarily.
\State \textbf{for} $i=0:N-1$
\State Sample $\bb u\sim$ Uniform[0,1]
\State Sample $\bb X^{\ast}\sim q(\bb X^{\ast}|\bb X^{(i)})=\mc N(\bb X^{(i)}+\tau\nabla\log\pi(\bb X^{(i)}),2\tau \mathbb{I})$
\State \textbf{if} {$\bb u<\min\{1,\dfrac{\pi(\bb X^{\ast})q(\bb X^{(i)}|\bb X^{\ast})}{\pi(\bb X^{(i)})q(\bb X^{\ast}|\bb X^{(i)})}\}$} \textbf{Then}
\State \hspace*{20pt}$\bb X^{(i+1)}\gets\bb X^{\ast}$
\State \textbf{else}
\State \hspace*{20pt} $\bb X^{(i+1)}\gets\bb X^{(i)}$
\State \textbf{endif}
\State \textbf{endfor}
\State \textbf{return} $\bb X^{(0)},\ldots,\bb X^{(N-1)}$
\end{algorithmic}
\end{algorithm}

We denote random variables with small bold-faced letters, random vectors by capital bold-faced letters, and their corresponding realizations by non-bold-faced letter. The MAL algorithm is outlined in Algorithm \ref{MAL_alg} wherein $\pi(\bb X)$ denotes the target probability distribution, $\tau$ is a positive real-valued parameter specifying the time-step used in the Euler-Maruyama approximation of the underlying Langevin dynamics, $N$ denotes the number of samples generated by the MAL algorithm, $q$ denotes the proposal distribution (a.k.a. transition kernel), $\mc N(\mu,\Sigma)$ denotes the multivariate normal distribution with mean vector $\mu$ and covariance matrix $\Sigma$, and, finally, $\mathbb{I}$ denotes the identity matrix. The sequence of samples generated by the MAL algorithm, $\bb X^{(0)},\bb X^{(1)}, \ldots$, is guaranteed to converge in distribution to $\pi(\bb X)$ \cite{robert2013monte}. It is worth noting that work in theoretical neuroscience has shown that MAL, outlined in Algorithm \ref{MAL_alg}, could be implemented in a neurally-plausible manner \cite{savin2014spatio,moreno2011bayesian}. In the following section, we propose a target distribution  $\pi(\bb X)$, allowing CCNNs to generate samples from a category of interest.

\section{The Proposed Framework}
\label{sec_proposed_framework}
In what follows, we propose a framework which transforms CCNNs into probabilistic generative models, thereby enabling them to generate samples from a category of interest. The proposed framework is based on the MAL algorithm given in Sec. \ref{sec_MAL_alg}. Let $f(X;W^\ast)$ denote the input-output mapping learned by a CCNN at the end of the training phase, and $W^\ast$ denote the set of weights for a CCNN after training.\footnote{Formally, $f(\cdot;W^\ast): \prod_{i=1}^n D_i\rightarrow \prod_{j=1}^m R_j$ where $D_i$ and $R_j$ denote the set of values that input unit $i$ and output unit $j$ can take on, respectively.} Upon termination of training, presented with input $X$, a CCNN outputs $f(X;W^\ast)$. Note that, in case a CCNN possesses multiple output units, the mapping $f(X;W^\ast)$ will be a vector rather than a scalar. To convert a CCNN into a probabilistic generative model, we propose to use the MAL algorithm with its target distribution $\pi(\bb X)$ being set as follows: 
\begin{eqnarray}
\label{eq_proposed}
\tilde{\pi}(\bb X)&\triangleq& p(\bb X|\bb Y=L_j)\nonumber \\
&=&\dfrac{1}{Z}\exp(-\beta||L_j-f(\bb X;W^\ast)||_2^2),
\end{eqnarray}
where $||\cdot||_2$ denotes the $l_2$-norm, $\beta\in\mathbb{R}_+$ is a \emph{damping factor}, $Z$ is the normalizing constant, and $L_j$ is a vector whose element corresponding to the desired class is $+0.5$  (i.e., its $j^{\text{th}}$ element) and the rest of its elements are $-0.5$s. The intuition behind Eq. (\ref{eq_proposed}) can be articulated as follows: {For an input instance $\bb X=X$ belonging to the desired class $j$,\footnote{In {counterfactual} terms, this is equivalent to saying: Had input instance $X$ been presented to the network, it would have classified $X$ in class $j$.} the output of the network $f(X;W^\ast)$ is expected to be close to $L_j$ in $l_2$-norm sense; in this light, Eq. (\ref{eq_proposed}) is adjusting the likelihood of input instance $X$ to be inversely proportional to the exponent of the said $l_2$ distance.}

For the reader familiar with probabilistic graphical models, the expression in Eq. (\ref{eq_proposed}) looks similar to the expression for the joint probability distribution of Markov random fields and probabilistic energy-based models, e.g., Restricted Boltzman Machines and Deep Boltzman Machines. However, there is a crucial distinction: The normalizing constant $Z$, the computation of which is intractable in general, renders learning in  those models computationally intractable.\footnote{{More specifically, $Z$ renders the computation of the gradient of the log-likelihood for those models intractable.}} The appropriate way to interpret Eq. (\ref{eq_proposed}) is to see it as a Gibbs distribution for a non-probabilistic energy-based model whose energy is defined as the square of the prediction error \cite{lecun2006tutorial}. Section 1.3 of \cite{lecun2006tutorial} discusses the topic of Gibbs distribution for non-probabilistic energy-based models in the context of discriminitive learning,  computationally modeled by $p(\bb Y|\bb X)$ (i.e., to predict a class given an input), and raises the same issue that we highlighted above regarding the intractability of computing the normalizing constant $Z$ in general. {In sharp contrast to \cite{lecun2006tutorial}, our framework is proposed for the purpose of {generating} examples from a desired class, as evidenced by Eq. (\ref{eq_proposed}) being defined in terms of $p(\bb X|\bb Y)$.} Also crucially, the intractability of computing $Z$ raises no issue for our proposed framework due to an intriguing property of the MAL algorithm according to which the normalizing constant $Z$ need not be computed at all.\footnote{The MAL algorithm inherits this property from the Metropolis-Hasting algorithm, which it uses as a subroutine.}

Due to Line 4 of Algorithm \ref{MAL_alg}, MAL's proposal distribution, $q$, requires the computation of $\nabla\log\tilde{\pi}(\bb X^{(i)})$, which essentially involves the computation of $\nabla f(\bb X^{(i)};W^\ast)$ (note that the gradient is operating on $\bb X^{(i)}$, and $W^\ast$ is merely treated as a set of {fixed} parameters). The multi-layer structure of CCNN ensures that $\nabla f(\bb X^{(i)};W^\ast)$ can be efficiently computed using Backpropagation. Alternatively, in settings where CCNNs recruit a small number of input units (hence, the cardinality of $\bb X^{(i)}$ is small), $\nabla f(\bb X^{(i)};W^\ast)$ can be obtained by introducing negligible perturbation to a component of input signal $\bb X^{(i)}$, dividing the resulting change in network's outputs by the introduced perturbation, and repeating this process for all components of input signal $\bb X^{(i)}$. It is worth noting that although the idea of computing gradients through introducing small perturbations would lead to a computationally inefficient approach for \emph{learning} CCNNs, it leads to a computationally efficient approach for \emph{generation}, as the number of input units are typically much fewer than the number of weights in CCNNs (and artificial neural networks, in general). It is also crucial to note that the normalizing constant $Z$ plays no role in the computation of $\nabla\log\tilde{\pi}(\bb X^{(i)})$.

\section{Simulations}
In this section we demonstrate the efficacy of our proposed framework through simulations. We particularly focus on learning which can be accomplished by two input and one output units. This permits visualization of the input-output space, which lies in $\mathbb{R}^3$. Note that our proposed framework can handle arbitrary number of input and output units; this restriction is solely for  ease of visualization. 

\subsection{Continuous-XOR Problem}
\label{sec_cont_xor}
\begin{figure}[h!]
\centering
\includegraphics[trim = 20mm 85mm 10mm 91mm, clip, width=4.4cm]{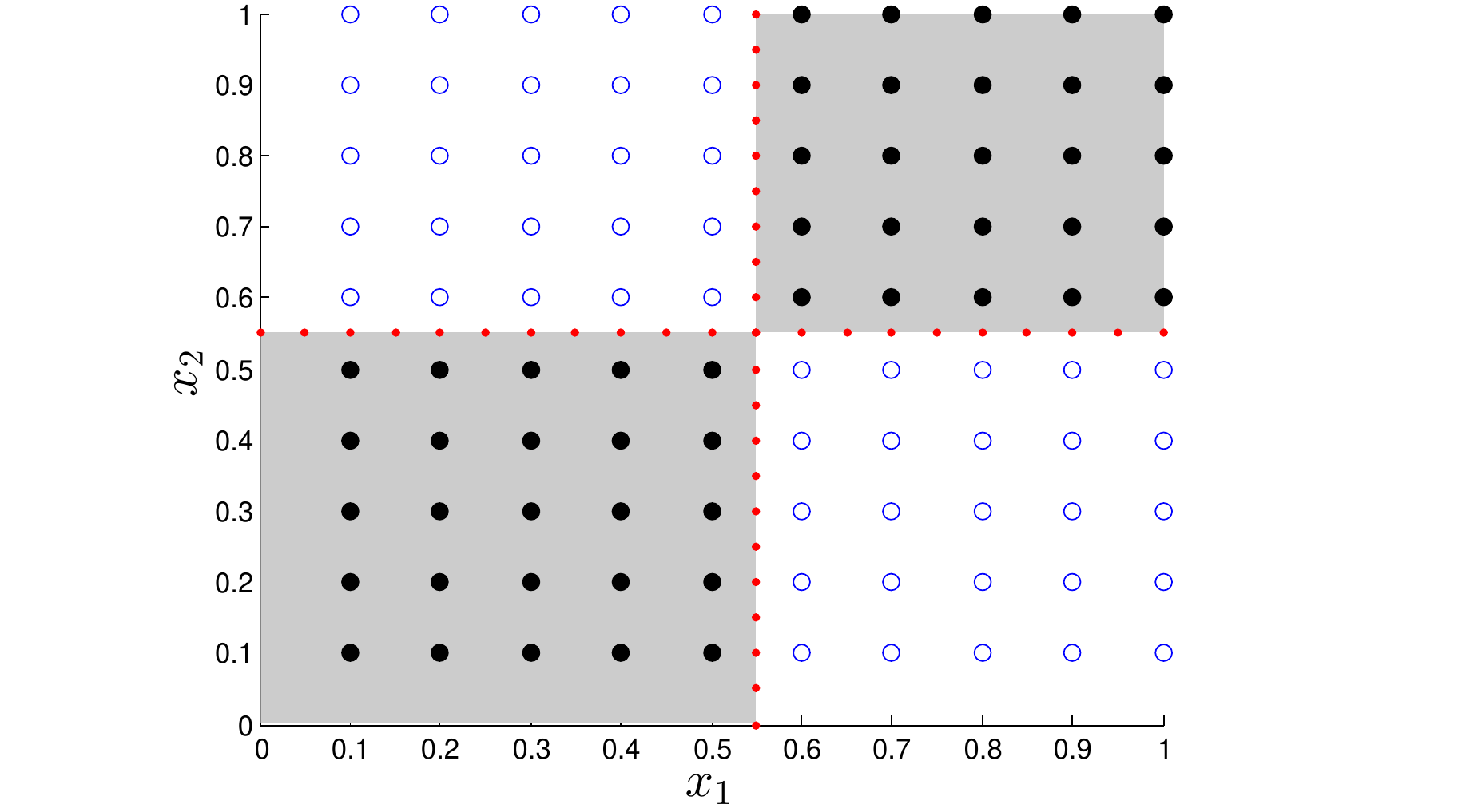}
\includegraphics[trim = 10mm 65mm 10mm 65mm, clip, width=4cm]{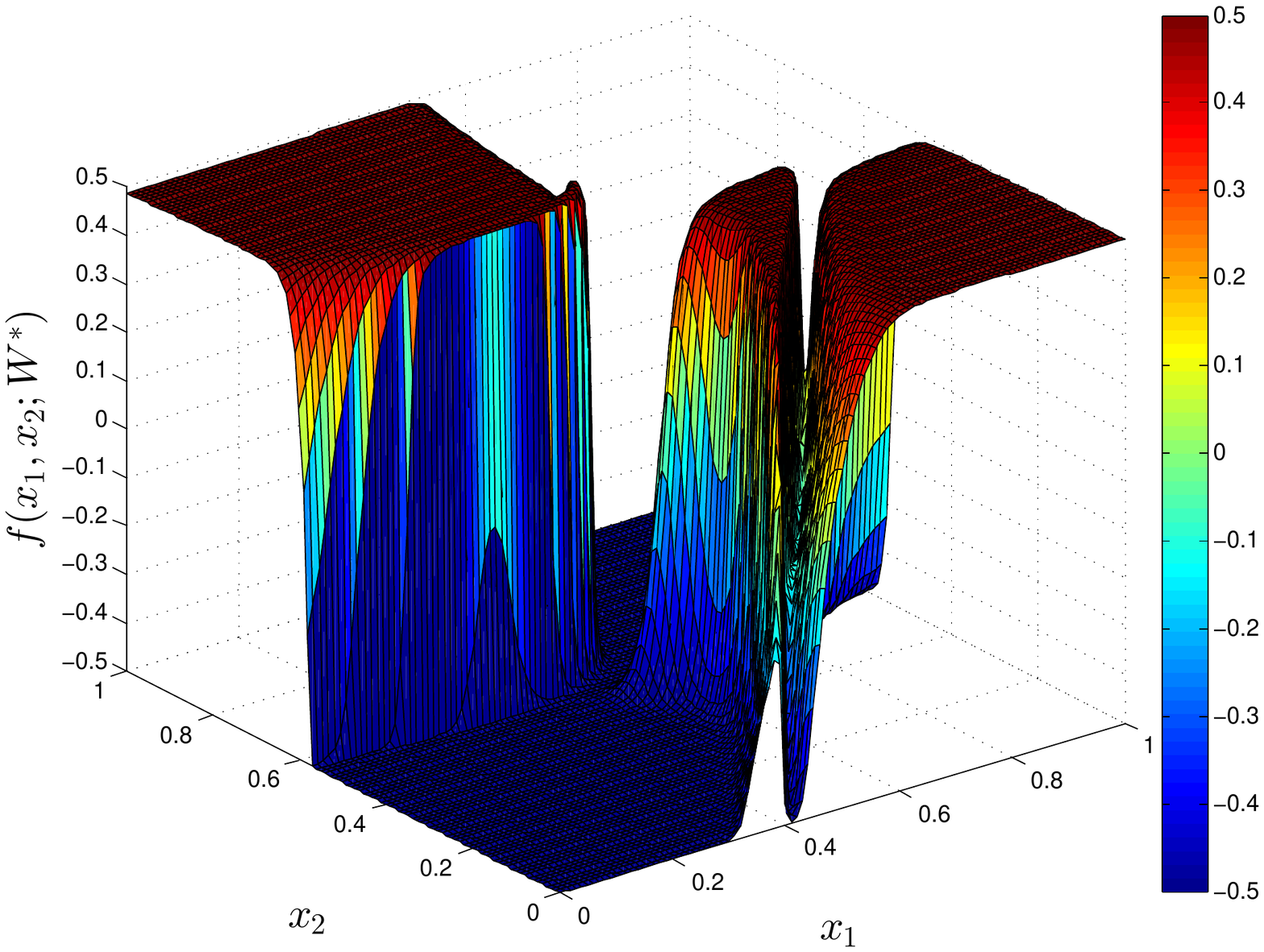}
\includegraphics[trim = 40mm 90mm 50mm 90mm, clip, width=5.8cm]{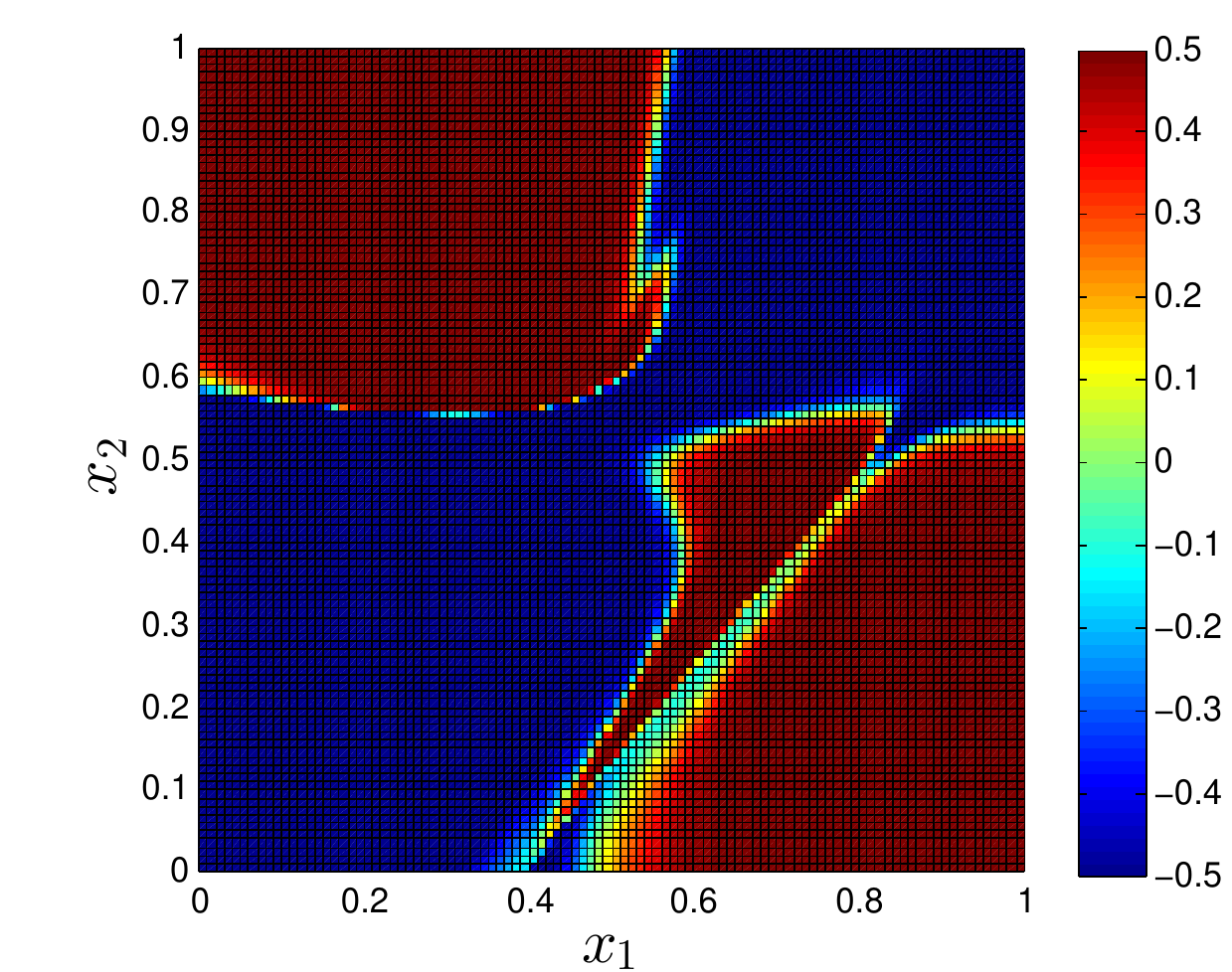}
\caption{A CCNN trained on the continuous-XOR classification task. Top-left: Training patterns. All the patterns in the gray quadrants are negative examples with label $-0.5$, and all the patterns in the white quadrants are positive examples with label $+0.5$. Red dotted lines depict the boundaries. Top-right: The input-output mapping, $f(x_1,x_2;W^\ast)$, learned by a CCNN, along with a colorbar. Bottom: The top-down view of the curve depicted in top-right, along with a colorbar.}
\label{fig_cont_xor_training_pattern}
\end{figure}

In this subsection, we show how our proposed framework allows a CCNN, trained on the continuous-XOR classification task (see Fig. \ref{fig_cont_xor_training_pattern}), to generate examples from a category of interest. The output unit has a symmetric sigmoidal activation function with range $-0.5$ and $+0.5$. The training set consists of $100$ samples in the unit-square $[0,1]^2$, paired with their corresponding labels. More specifically, the training set is comprised of all the ordered-pairs starting from $(0.1,0.1)$ and going up to $(1,1)$ with equal steps of size $0.1$, paired with their corresponding labels (i.e., $+0.5$ for positive samples and $-0.5$ for negative samples); see Fig. \ref{fig_cont_xor_training_pattern}(top-left). After training, a CCNN with $6$ hidden layers is obtained whose input-output mapping, $f(x_1,x_2;W^\ast)$, is depicted in Fig. \ref{fig_cont_xor_training_pattern}(top-right).\footnote{Due to the inherent randomness in CCNN construction, training could lead to networks with different structures. However, since in this work we are solely concerned with {generating} examples using CCNNs rather than how well CCNNs could learn a given discriminitive task, we arbitrarily pick a learned network. Note that our proposed framework can handle CCNNs with arbitrary structures; in that light, the choice of network is without loss of generality.}

\begin{figure*}[t!]
    \centering
    \subfloat[$N=2000$, $AR=99.55\%$]{{\includegraphics[trim = 12mm 65mm 10mm 58mm, clip,width=5.5cm]{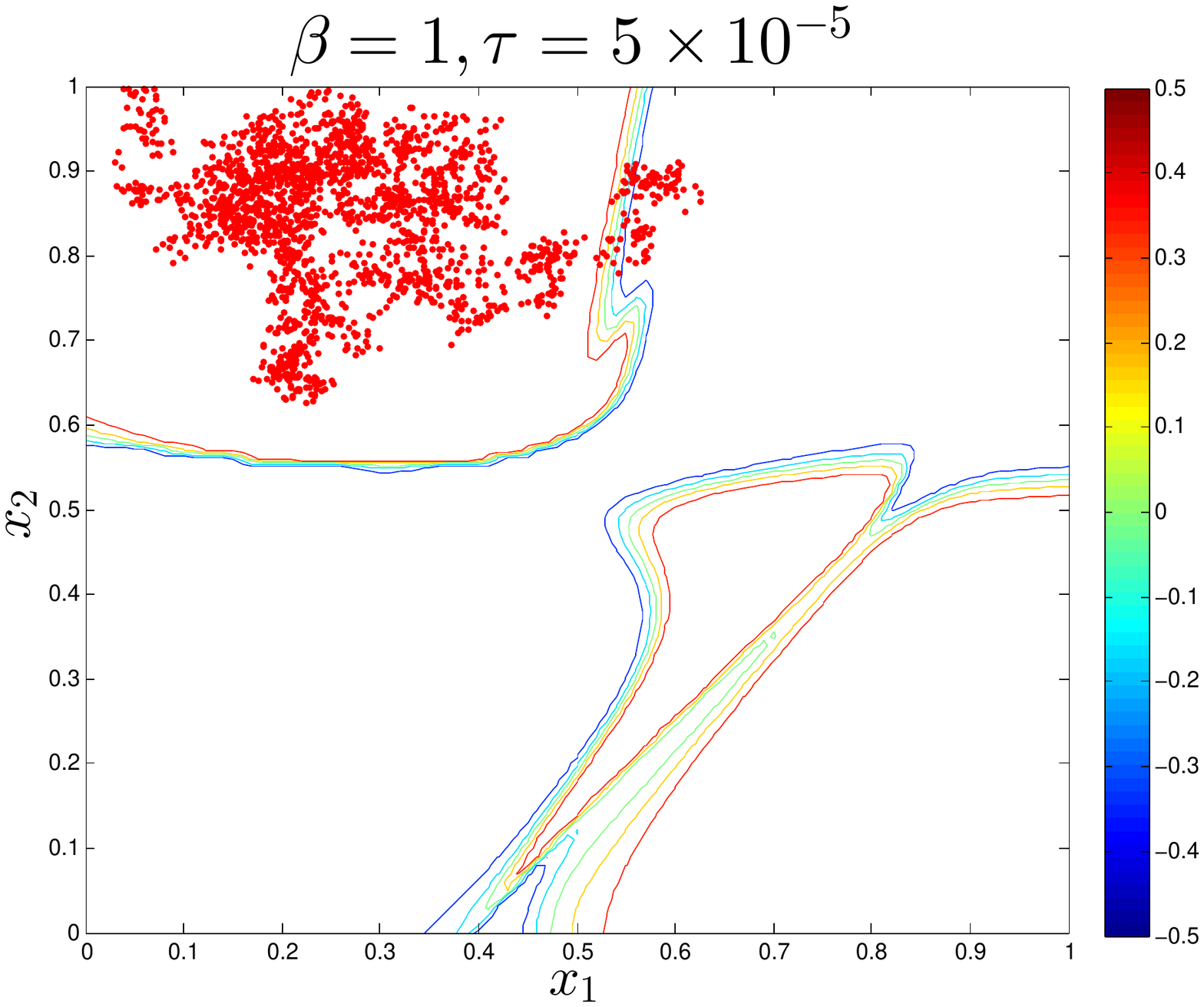} }}
    \hspace*{1pt}
    \subfloat[$N=2000$, $AR=75.25\%$]{{\includegraphics[trim = 12mm 65mm 10mm 58mm, clip,width=5.5cm]{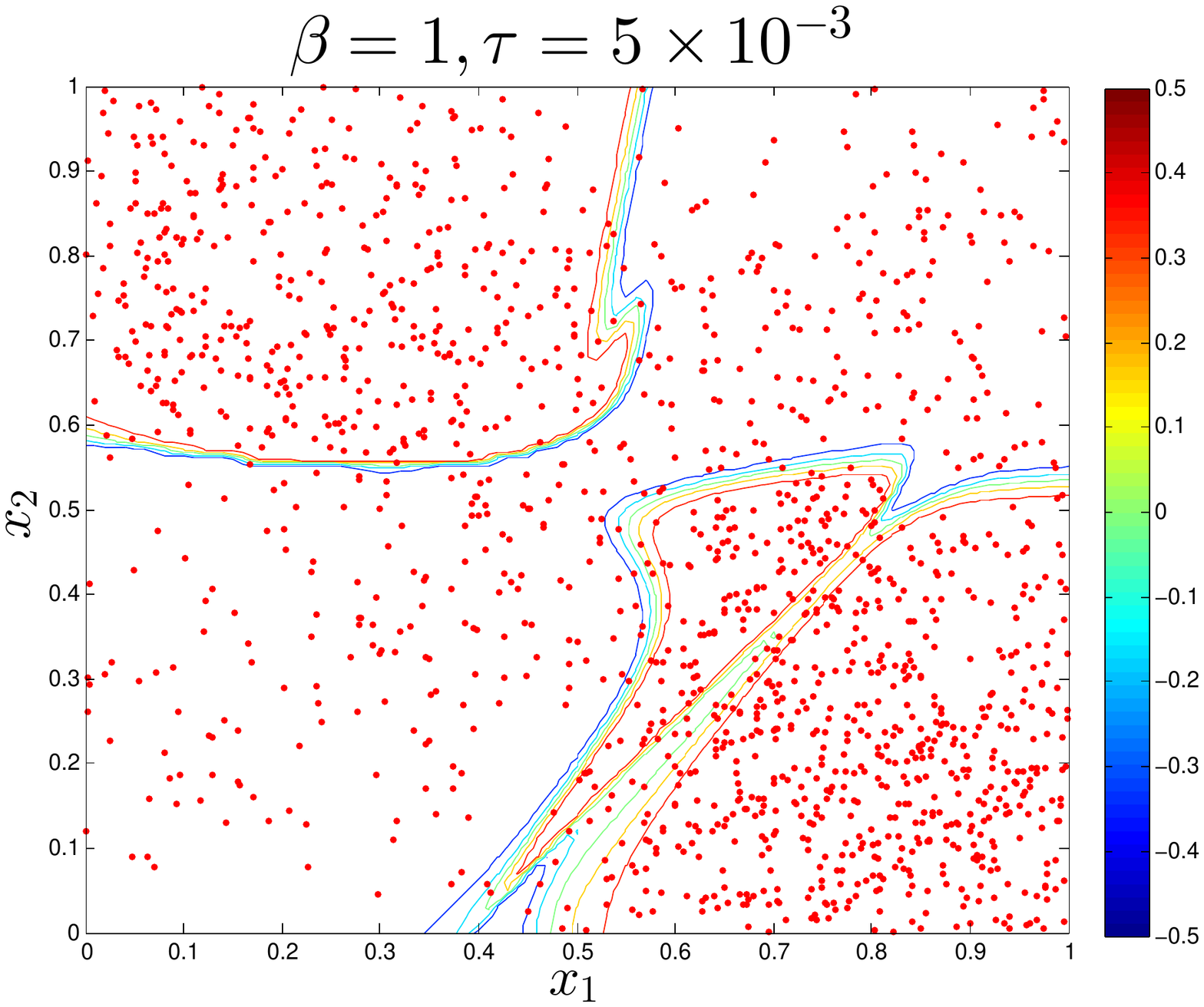} }}
    \hspace*{1pt}
    \subfloat[$N=2000$, $AR=57.85\%$]{{\includegraphics[trim = 12mm 65mm 10mm 58mm, clip,width=5.5cm]{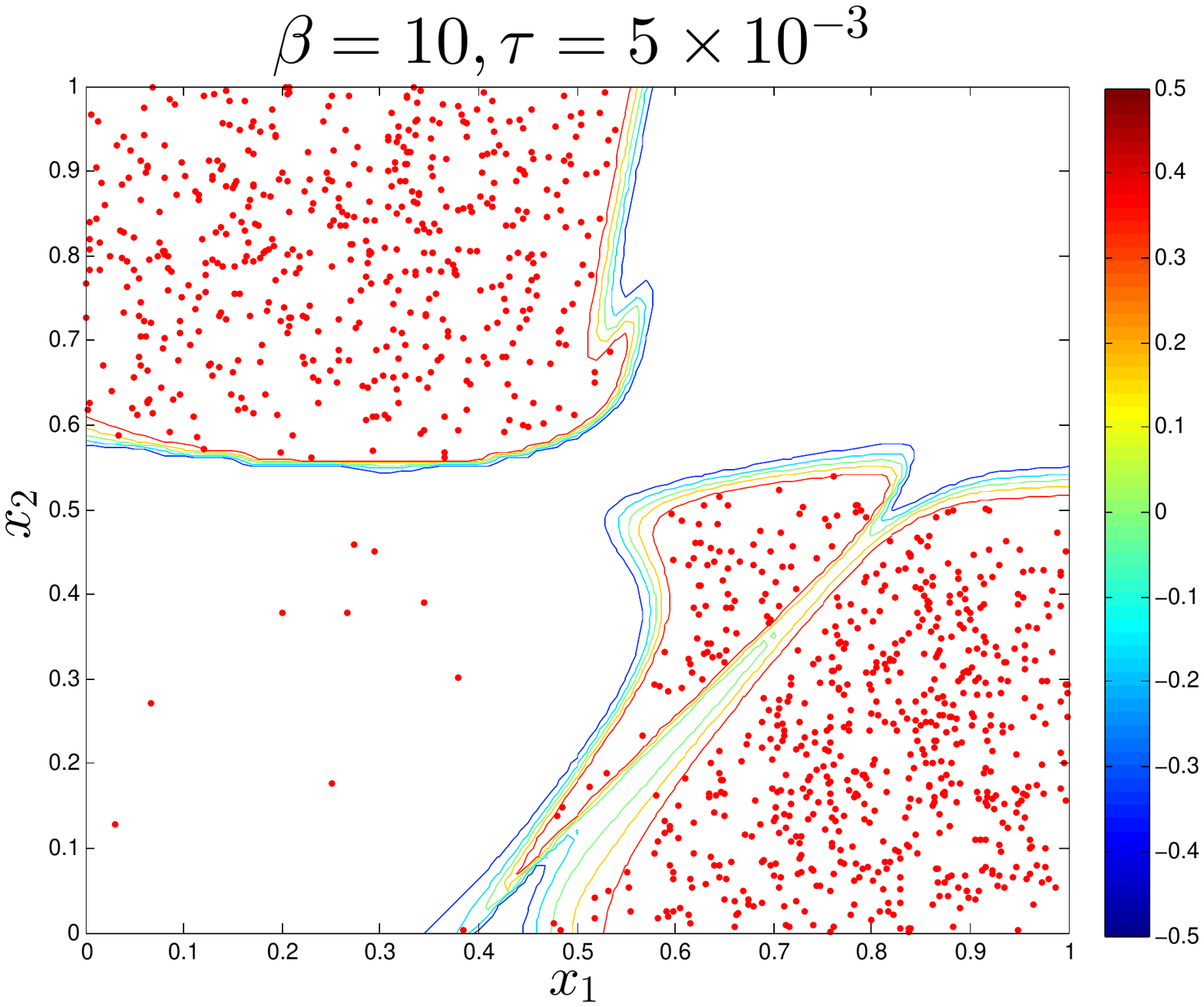} }}
    \caption{Generating example for the positive category, under various choices for MAL parameter $\tau$ and damping factor $\beta$. Contour-plot of the learned mapping, $f(x_1,x_2;W^\ast)$, along with its corresponding colorbar is shown in each sub-figure. Generated samples are depicted by red dots. $N$ denotes the total number of samples generated by MAL, and $AR$ denotes the corresponding acceptance rate. \textbf{(a)} $\tau=5\times 10^{-5}$ leads to a very slow exploration of the input space. \textbf{(b)} $\tau=5\times 10^{-3}$ leads to an adequate exploration of the input space, however, $\beta=1$ is not penalizing undesirable input regions severely enough. \textbf{(c)} A desirable performance is achieved by $\tau=5\times 10^{-3}$ and $\beta=10$.}
\label{fig_xor_beta_tau}
\end{figure*}

Fig. \ref{fig_xor_beta_tau} shows the efficacy of our proposed framework in enabling CCNNs to generate samples from a category of interest, under various choices for MAL parameter $\tau$ (see Algorithm \ref{MAL_alg}) and damping factor $\beta$ (see Eq. (\ref{eq_proposed})); generated samples are depicted by red dots. For the results shown in Fig. \ref{fig_xor_beta_tau}, the category of interest is the category of positive examples, i.e., the category of input patterns which, upon being presented to the (learned) network, would be classified as positive by the network. Because $\tau$ controls the amount of jump between consecutive proposals made by MAL, the following behavior is expected: For small $\tau$ (Fig. \ref{fig_xor_beta_tau}(a)) consecutive proposals are very close to one another, leading to a slow exploration of the input domain. As $\tau$ increases, bigger jumps are made by MAL (Fig. \ref{fig_xor_beta_tau}(b)).\footnote{Yet, too large a $\beta$ is not good either, leading to a sparse and coarse-grained exploration of the input space. Some measures have been proposed in computational statistics for properly choosing $\tau$; cf. \cite{roberts1998optimal}.} Parameter $\beta$ controls how severely deviations from the desired class label (here, $+0.5$) should be penalized. The larger the parameter $\beta$, the more severely such deviations are penalized and the less likely it becomes for MAL to make moves toward such regions of input space. Acceptance Rate (AR), defined as the number of accepted moves divided by the total number of suggested moves, is also presented for the results shown in Fig. \ref{fig_xor_beta_tau}. Fig. \ref{fig_xor_beta_tau}(c) shows that for $\tau=5\times 10^{-3}$ and $\beta=10$, our proposed framework demonstrates a desirable performance: virtually all of the generated samples fall within the desired input regions (i.e., the regions associated with hot colors, signaling the closeness of network's output to $+0.5$ in those regions; see Fig. \ref{fig_cont_xor_training_pattern}(bottom)) and the desired regions are adequately explored (i.e., all hot-colored input regions being visited and almost evenly explored). 

\begin{figure}[h!]
    \centering
    \subfloat{{\includegraphics[trim = 12mm 65mm 10mm 58mm, clip,width=7cm]{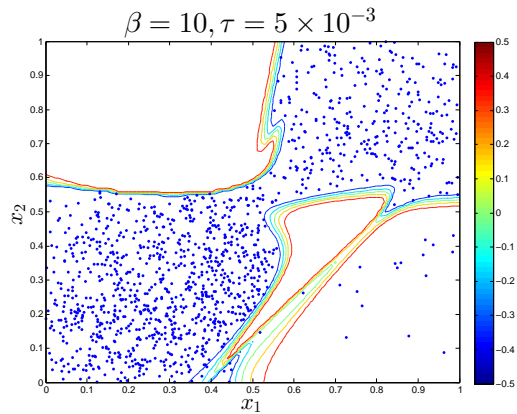} }}
    \caption{Generating example for the negative category, with $\tau=5\times10^{-3}, \beta=10$. Generated samples are shown by blue dots. Total number of samples generated is $N=2000$, with $AR=65.13\%$.}
\label{fig_xor_beta_tau_negative}
\end{figure}

Results shown in Fig. \ref{fig_xor_beta_tau} depict all the first $N=2000$ samples generated by MAL, without excluding the so-called {burn-in} period. In that light, the result shown in Fig. \ref{fig_xor_beta_tau}(c) nicely demonstrates how MAL---by directing its suggestions toward the direction of gradient and therefore making moves toward regions with high likelihood---could alleviate the need for discarding a (potentially large) number of samples generated at the beginning of an MCMC which are assumed to be unrepresentative of equilibrium state, a.k.a. the burn-in period. Fig. \ref{fig_xor_beta_tau_negative} shows the performance of our framework in enabling the learned CCNN to generate from the category of negative examples, with $\tau=5\times10^{-3}$ and $\beta=10$.   

\subsection{Two-Spirals Problem}
\begin{figure}[h!]
\centering
\includegraphics[trim = 15mm 60mm 10mm 75mm, clip, width=4cm]{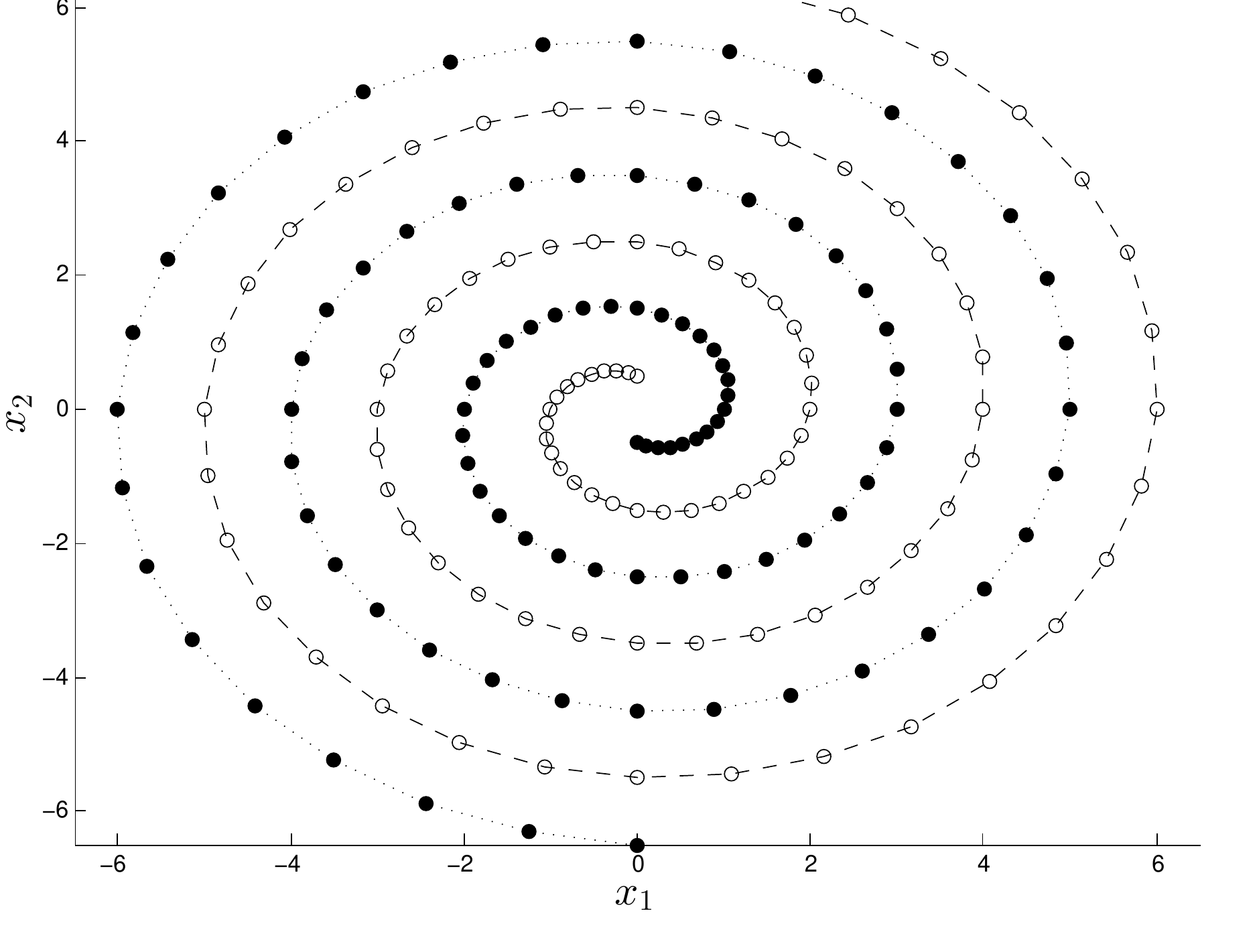}
\includegraphics[trim = 10mm 65mm 10mm 65mm, clip, width=4.4cm]{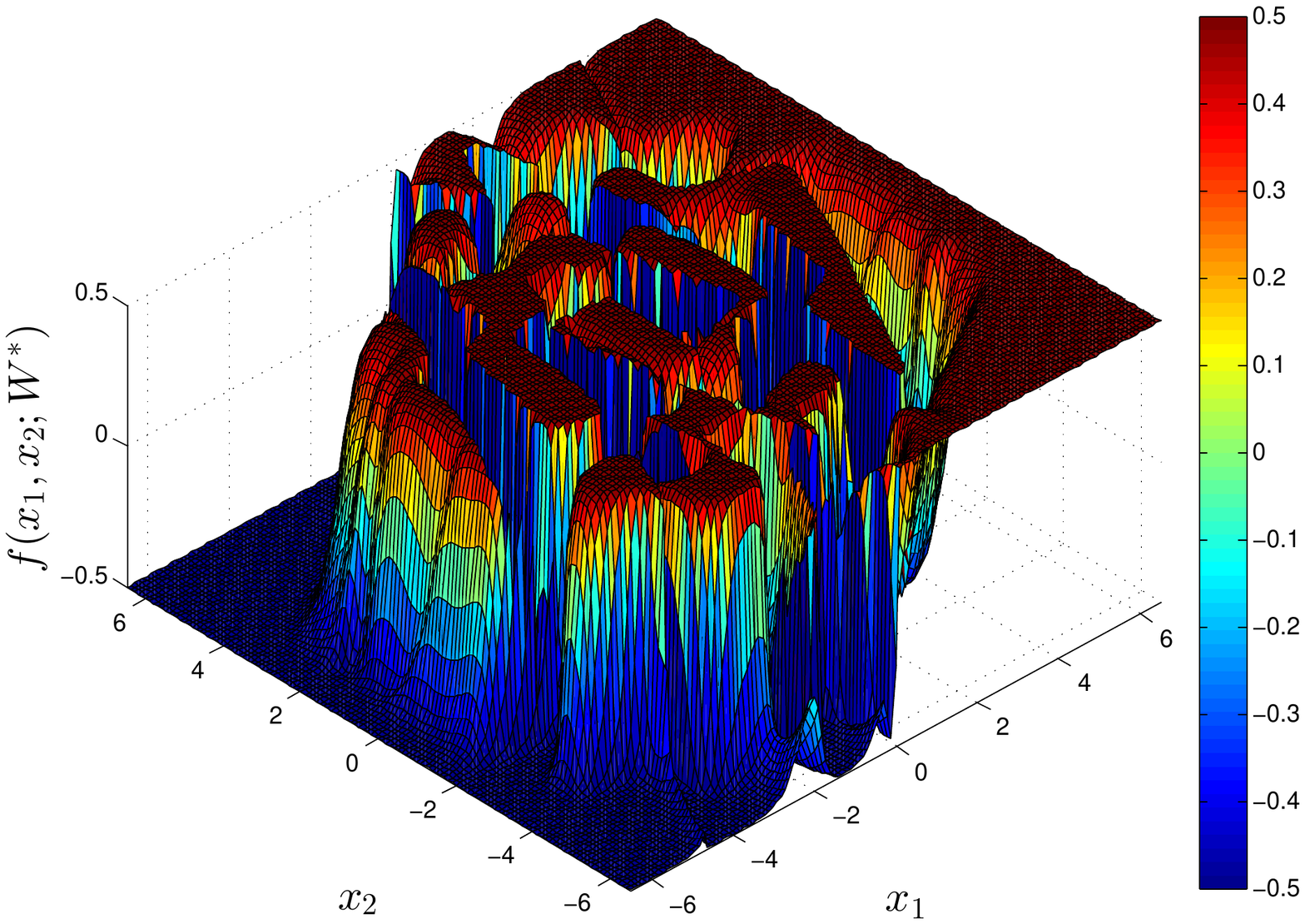}
\includegraphics[trim = 40mm 85mm 45mm 90mm, clip, width=5.8cm]{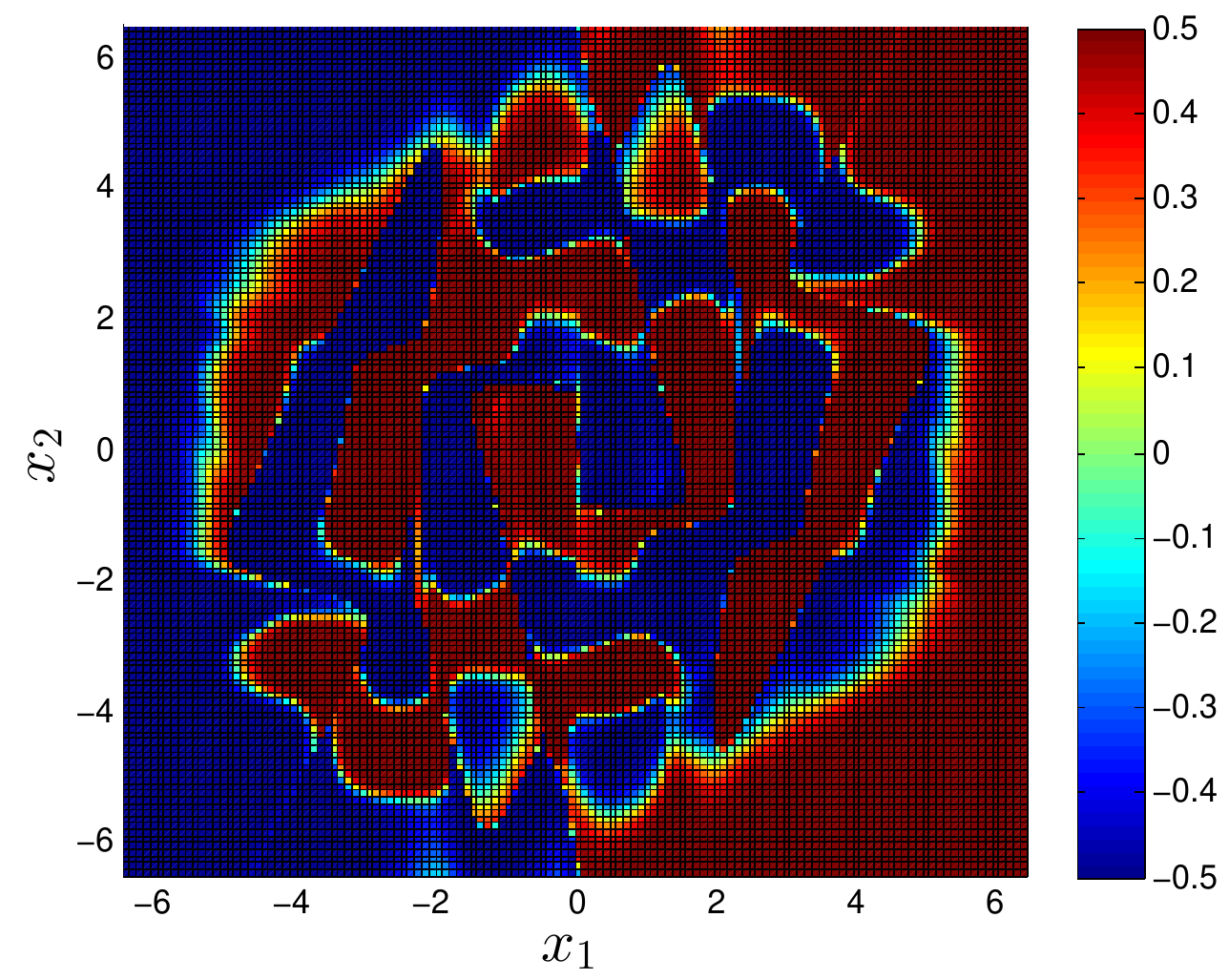}
\caption{A CCNN trained on the two-spirals classification task. Top-left: Training patterns. Positive patterns (associated with label $+0.5$) are shown by hollow circles, and negative patterns (associated with label $-0.5$) by black circles. Positive spiral is depicted by a dashed line, and negative spiral by a dotted line. Top-right: The input-output mapping, $f(x_1,x_2;W^\ast)$, learned by a CCNN, along with a colorbar. Bottom: The top-down view of the curve depicted in top-right, along with a colorbar.}
\label{fig_two_spirals_training_pattern}
\end{figure}

Next, we show how our proposed framework allows a CCNN, trained on the famously difficult Two-Spirals classification task (Fig. \ref{fig_two_spirals_training_pattern}), to generate examples from a category of interest. The output unit has a symmetric sigmoidal activation function with range $-0.5$ and $+0.5$. The training set consists of $194$ samples ($97$ samples per spiral), in the square $[-6.5,6.5]^2$, paired with their corresponding labels ($+0.5$ and $-0.5$ for positive and negative samples, respectively). The training pattern is shown in Fig. \ref{fig_two_spirals_training_pattern}(top-left); cf.  \cite{chalup2007variations} for details. After training, a CCNN with $14$ hidden layers is obtained whose input-output mapping, $f(x_1,x_2;W^\ast)$, is depicted in Fig. \ref{fig_two_spirals_training_pattern}(top-right).

\begin{figure}[t!]
\centering
\includegraphics[trim = 10mm 60mm 10mm 60mm, clip, width=6.5cm]{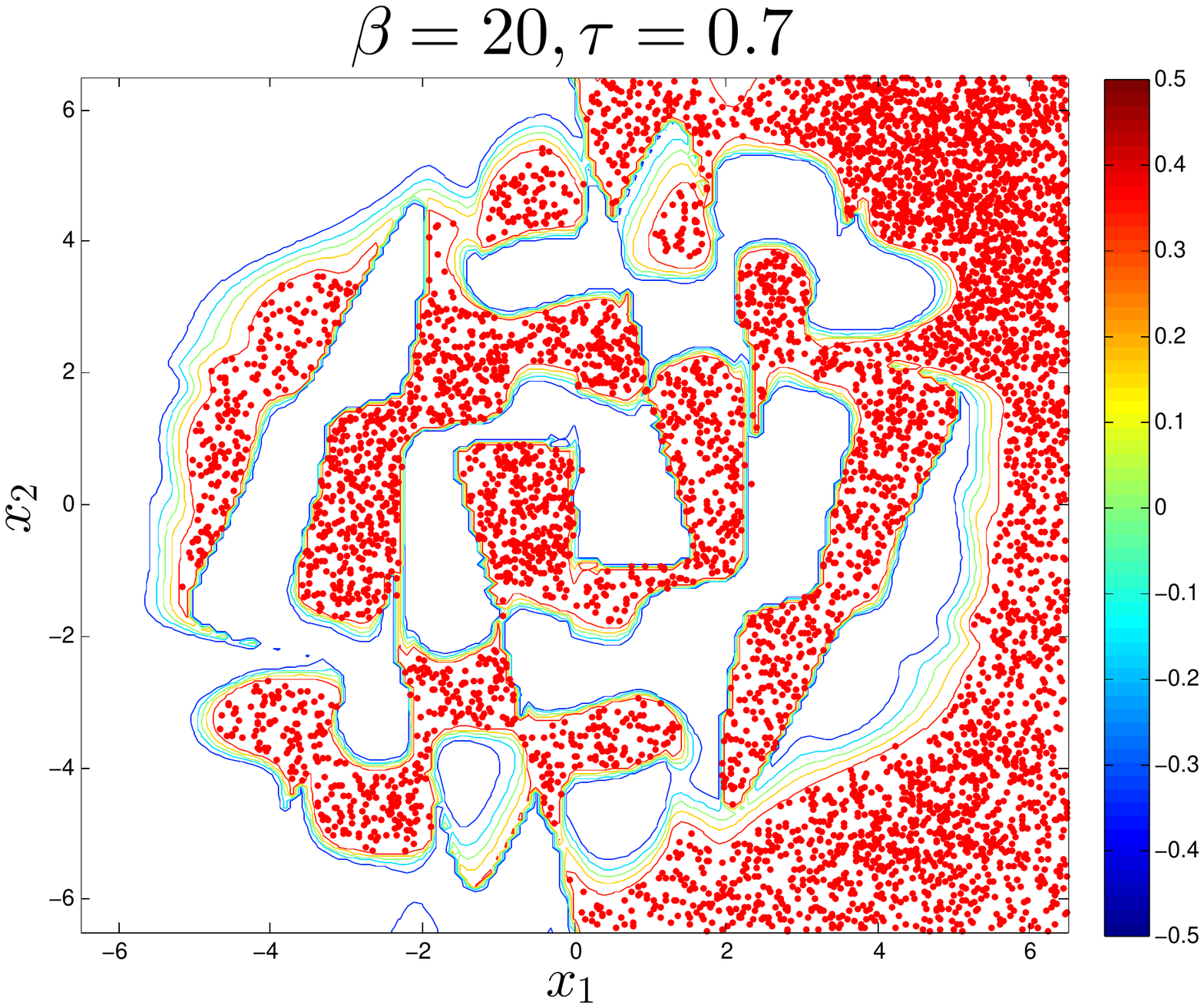}
\includegraphics[trim = 10mm 65mm 10mm 60mm, clip, width=6.5cm]{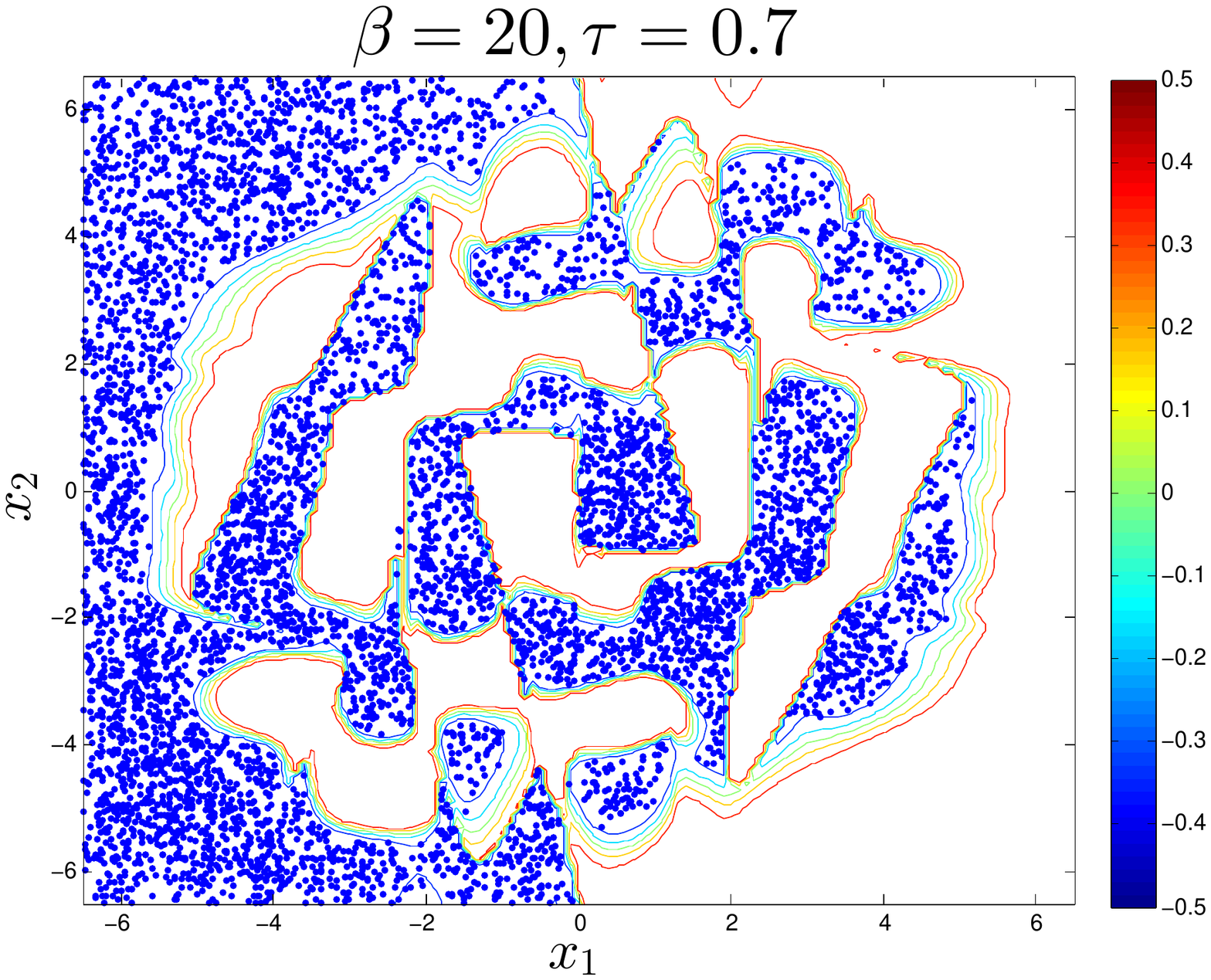}
\caption{Generating example for the positive and negative categories, with $\beta=20$ and $\tau=0.7$.  Contour-plot of the learned mapping, $f(x_1,x_2;W^\ast)$, along with its corresponding colorbar is shown in each sub-figure. $N$ denotes the total number of samples generated by MAL, and $AR$ denotes the corresponding acceptance rate. Top: Generating example for the positive category, with $N=15000$ and $AR=40.69\%$. Generated samples are depicted by red dots. Bottom: Generating example for the negative category, with $N=15000$ and $AR=40.28\%$. Generated samples are depicted by blue dots.}
\label{fig_two_spirals_positive_negative}
\end{figure}

Fig. \ref{fig_two_spirals_positive_negative}(top) and Fig. \ref{fig_two_spirals_positive_negative}(bottom) show the efficacy of our proposed framework in enabling CCNNs to generate samples from the positive and negative categories, respectively. Although similar patterns of behavior observed in Sec. \ref{sec_cont_xor} due to increasing/decreasing $\beta$ and $\tau$ are observed here as well, due to the lack of space such results are omitted. Note that the results shown in Fig. \ref{fig_two_spirals_positive_negative} depicts all the first $N=15000$ samples generated by MAL, without excluding the burn-in period. In that light, the results shown in Fig. \ref{fig_two_spirals_positive_negative}(top) and Fig. \ref{fig_two_spirals_positive_negative}(bottom) demonstrate once again the efficacy of MAL in alleviating the need for discarding a (potentially large) number samples generated at the beginning of an MCMC run. 

\begin{figure}[t!]
\centering
\includegraphics[trim = 10mm 60mm 10mm 57mm, clip, width=6.29cm]{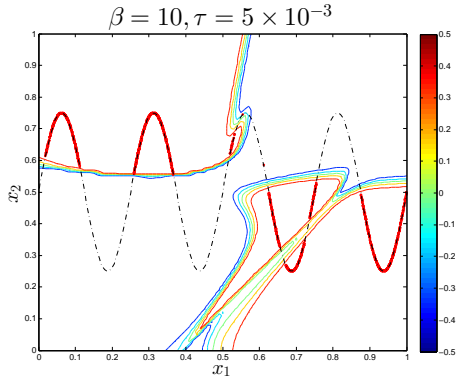}
\caption{Generating example for the positive category, under constraint $x_2=0.25\sin(8\pi x_1)+0.5$ (dashed dotted curve), with $N=5000$ and $AR=39.82\%$. Contour-plot of the learned mapping, $f(x_1,x_2;W^\ast)$, along with its corresponding colorbar is depicted. Generated samples are shown by red dots (due to high density, individual samples may not be easily visible).}
\label{fig_cont_xor_constrained}
\end{figure}

Interestingly, our proposed framework also allows CCNNs to generate samples subject to some forms of constraints. For example, Fig. \ref{fig_cont_xor_constrained} demonstrates how our proposed framework enables a CCNN, trained on the continuous-XOR classification task (see Sec. \ref{sec_cont_xor}), to generate examples from the positive category, under the following constraint: Generated samples must lie on the curve $x_2=0.25\sin(8\pi x_1)+0.5$. To generate samples from the positive category while satisfying the said constraint, MAL adopts our proposed target distribution given in Eq. (\ref{eq_proposed}), and treats $x_1$ as an independent and $x_2$ as a dependent variable.

\section{General Discussion}
Although we discussed our proposed framework in the context of CCNNs, it can be straightforwardly extended to handle other kinds of artificial neural networks, e.g., multi-layer perceptron and convolutional neural networks. Furthermore, our proposed framework, together with recent work in theoretical neuroscience showing possible neurally-plausible implementations of MAL \cite{savin2014spatio,moreno2011bayesian}, suggests an intriguing {modular} hypothesis according to which generation could result from two separate modules interacting with each other (in our case, a CCNN and a neural network implementing MAL). This hypothesis  yields the following prediction: There should be some brain impairments which lead to a marked decline in a subject's performance in generative tasks (i.e., tasks involving imagery, or imaginative tasks in general) but leave the subject's learning abilities (nearly) intact. Studies on learning and imaginative abilities of hippocampal amnesic patients already provide some supporting evidence for this idea  \cite{hassabis2007patients,spiers2001hippocampal,brooks1976can}.

{According to Line 4 of Algorithm 1, to generate the $i^{\text{th}}$ sample, MAL requires to have access to a fine-tuned, Gaussian noise with mean $\bb X^{(i)}+\tau\nabla\log\pi(\bb X^{(i)})$ for its proposal distribution $q$. Recently \citeA{savin2014spatio} showed how a network of leaky integrate-and-fire neurons could implement MAL in a neurally-plausible manner. However, as \citeA{gershman2016complex} point out, Savin and Deneve leave unanswered what the source of that fine-tuned Gaussian noise could be. Our proposed framework may provide an explanation, not for the source of Gaussian noise, but for its fine-tuned mean value. According to our modular account, the main component of the mean value, which is $\nabla\log\pi(\bb X^{(i)})$, may come from another module (in our case a CCNN) which has learned some input-output mapping $f(X;W^\ast)$, based on which the target distribution $\pi(\bb X^{(i)})$ is defined (see Eq. (1)).}     

The idea of sample generation under constraints could be an interesting line of future work. Humans clearly have the capacity to engage in imaginative tasks under a variety of constraints, e.g., when given incomplete sentences or fragments of a picture people can generate possible completions; cf. \cite{sanborn2016bayesian}. Also, our proposed framework can be used to let a CCNN generate samples from a category of interest at any stage during CCNN construction. In that light, our proposed framework, along with a neurally-plausible implementation of MAL, gives rise to a \emph{self-organized generative model}: a generative model possessing the self-constructive property of CCNNs. Such self-organized generative models could provide a wealth of developmental hypotheses as to how the imaginative capacities of children change over development, and models with quantitative predictions to compare against. We see our work as a step towards such models.

\section*{Acknowledgment}
This work is funded by an operating grant to TRS from the Natural Sciences and Engineering Research Council of Canada. 

\renewcommand\bibliographytypesize{\scriptsize}
\bibliographystyle{apacite}
\setlength{\bibleftmargin}{.125in}
\setlength{\bibindent}{-\bibleftmargin}
\bibliography{ref}
\end{document}